# MECHANISMS OF

# THE EQUILIBRIUM-RANGE FORMATION

# IN THE WIND-WAVE SPECTRUM.

# NUMERICAL SIMULATION

Polnikov V.G.

Obukhov Institute of Atmospheric Physics of RAS, Moscow, Russia 119017

e-mail: Polnikov@mail.ru

ABSTRACT

The issue of the equilibrium-range formation in the wind-wave spectrum is studied by a direct numerical simulation. The evolution equation of wind-wave spectrum is numerically solved with using an exact calculation of the Hasselmann kinetic integral and involving various modifications of known parametrizations for the wave-pumping mechanism due to wind and the wave-dissipation mechanism due viscosity of the upper layer. It is shown that the balance of the last two mechanisms is responsible for a formation of stationary part of the wind-wave spectrum. This balance allows establishing any prescribed form of the equilibrium part of the spectrum, if the proper representations of input and dissipation mechanisms are chosen.

### 1. Motivation

The point of studying physics responsible for the equilibrium-range formation in the frequency spectrum of wind waves (shortly - "the tail" of spectrum) is one of traditional problems of the wind-wave theory. Its importance is stipulated by the fact that it is the balance of evolution mechanisms for wind-waves responsible for the spectrum-tail formation. Thus, the reliable measurement of the spectrum-tail shape and its theoretical interpretation allow us to judge a correctness of our understanding evolution mechanisms for wind-waves and formation of their spectrum.

The beginning of studying this issue was initiated over 60 years ago by Phillips (1958). It was based on a simple consideration of dimensions. Phillips has proposed the equilibrium range of the frequency spectrum for wind waves in the form

$$S(\sigma) = \alpha_F g^2 \sigma^{-5} \quad \text{(for } \sigma > \sigma_p\text{)} \tag{1}$$

which was later called as the "Phillips' spectrum". Here, $S$ is the energy frequency-spectrum for water surface elevations, given in units $L^2T$, $\sigma$ is the circular frequency (rad/s), $\sigma_p$ is the peak frequency corresponding to the peak of spectrum $S(\sigma)$; $\alpha_F$ is the so-called "Phillips' constant", the value of which is not essential here, and $g$ is the acceleration due to gravity. Justification of the Phillips' spectrum shape is solely based on the formal assumption that the balance of evolution mechanisms for wind waves in the equilibrium range of spectrum does not depend on the wind speed, U, causing these waves. This formula has no more content.

Form (1) has been repeatedly confirmed by numerous field observations, for example, (Pierson and Moskowitz, 1964; Hasselmann et al, 1973, among others), in which a stability of realization of form (1) was observed for various values of spatially homogeneous and constant wind. Herewith, in the project JONSWAP (Hasselmann et al, 1973), it was found that a magnitude of $\alpha_F$ depends on wave-formation conditions. More details about the Phillips' spectrum are not needed here.

In the 60-ies of the last century, some theoretical models dealing with the wind-wave spectrum tail were appeared (Kitaigorodskii 1962; Zakharov and Filonenko, 1966), predicting its shape of the form

$$S(\sigma) = \alpha_T u_* g \sigma^{-4} \propto \sigma^{-4} \qquad . \tag{2}$$

In this case, the intensity of the spectrum tail depends on the friction velocity, $u_*$. (As is known, the link $u_*$ with the wind at a standard horizon, e.g., $U_{10}$, is given by a simple linear relationship). Kitaigorodskii (1962) has found form (2) by means of dimensions consideration, assuming that the rear part of wind-wave spectrum is formed by the Kolmogorov's flux of energy toward upper frequencies, given the value $\varepsilon = u_* g$. However, it was only a hypothesis.

Zakharov and Filonenko (1966) have obtained the spectrum of form (2) as the exact stationary solution of the four-wave kinetic integral (Hasselmann, 1962), if the power-like spectrum takes place on the whole frequency band, from 0 to ∞, and spread isotopically over the angle. The authors identified form (2) as the Kolmogorov's spectrum of the constant energy flux to the upper frequencies. But, in contrast to the Kitaigorodskii approach, they specified the mechanism determining the flux, namely, the four-wave nonlinear interactions in waves.

Several years later, the experimental laboratory work of Toba (1972) was appeared, in which he has showed for the first time that the spectrum of form (2) is well observed in the tank experiments. Since this time, spectrum of form (2) was called as the "Toba's spectrum", and the Phillips' spectrum was revised in favor of the Kitaigorodskii-Zakharov-Toba results in a huge series of theoretical works, including (Kitaigorodskii et al, 1975; Kitaigorodskii, 1983; Zakharov and Zaslavskii, 1982; Phillips 1985; and others). In particular, developing an analytical approach to solving the kinetic integral, Zakharov and Zaslavskii (1982) have found another kind of Kolmogorov-type solution

$$S(\sigma) \propto \sigma^{-11/3}, \qquad (3)$$

corresponding to the wave-action flux towards the lower frequencies. The same result was established in (Kitaigorodskii, 1983) by means of dimensional consideration. Since this time, the idea that Kolmogorov-like forms (2) and (3) could realizes in the natural wind-waves was actively discussed. Last theoretical version of crucial importance of non-linear interactions in the spectrum-tail formation for wind waves was presented in the recent paper (Zakharov and Badulin, 2011). Details of this work will be discussed below.

Regarding to analysis of observations, the field measurements data of JONSWAP (Hasselmann et al, 1973) were even revised (Kahma and Calkoen, 1992) in favor of form (2). In addition to this, there were executed several rigorous measurements in open (but small) reservoirs (see, Donelan et al, 1985; and overview paper: The WISE group, 2007), confirming the existence of form (2) in field waves.

Note however that there is no any clear theoretical proof of existence a constant flux of energy (or wave action) through the wind-wave spectrum under the natural wave-generation conditions, when the input and dissipation of energy are spread through the whole frequency range of the spectrum. Namely, it seems doubtful a presence of the inertial frequency range, in which the nonlinear mechanism of evolution "works" alone.

Moreover, for a long time it is known (Polnikov, 1989), and it has been recently especially shown by numerical calculations with two algorithms (Polnikov and Uma, 2014) that no form of the spectrum, given on a limited frequency band, does put the kinetic integral to zero. Herewith, these conclusions are in consistent with the results found in other papers by the same author (Polnikov, 1994, 2001). There it was numerically shown that the Kolmogorov's spectra are actually appearing due to nonlinear interactions among waves, when the energy source and sink are separated on the frequency band, i.e. the inertial interval in the spectrum of waves does really take place.

At the same time, a special field-study of the equilibrium part of spectrum for the natural wind waves (Rodrigues and Soares, 1999) have shown that the decay law of the spectrum can vary within very wide limits (from $\sigma^{-3}$ to $\sigma^{-7}$). And these results have not a clear explanation in terms of nonlinear interactions.

Note that in most cases, the discussed issue was considered by theoreticians with analytical methods applied for cases of simple conditions for wave generation, assuming various simplifications, approximations and hypotheses. However, at present the numerical simulation of wind wave has already reached a level when its accuracy is close to one of direct buoy measurements (Samiksha et al, 2015). Sometimes such a simulation even exceeds the accuracy of direct satellite measurements (Kubryakov et al, 2016). The foregoing means a considerable progress in understanding physics of all the evolution mechanisms for wind waves. Therefore, taking into account an availability of accurate algorithms for the exact numerical solution of the nonlinear kinetic equation (Polnikov, 1990; van Vledder, 2005), the discussed question: what mechanisms are responsible for the formation of the spectrum tail for wind waves, can be easily solved by a direct numerical simulation of the spectrum evolution for wind waves. The present work is devoted to solving this task.

2. **The essence of the problem**

The problem to be solved is the following. It is well known that the evolution of the two-dimensional energy spectrum of wind waves, $S(\sigma, \theta; t, x, y)$, is described by the equation corresponding to the energy conservation of each wave-spectrum components, $S(\sigma, \theta)$, in the time-space coordinates, $(t, x, y)$. In the simplest case, it has the form (Komen et al, 1994)

$$\frac{dS}{dt} \equiv \frac{\partial S}{\partial t} + C_{gx}\frac{\partial S}{\partial x} + C_{gy}\frac{\partial S}{\partial y} = F \equiv In(S,U) + Nl(S) - Dis(S,U). \tag{4}$$

In equation (4), the left-hand-side means the total derivative of the spectrum with respect to time, and the right-hand-side is the so-called source function $F$ (hereinafter - $SF$). The left side of (4) is responsible for the mathematical part of the numerical model for wind waves, which is complemented by the boundary and initial conditions, and by the input wind-field. The $SF$ contains the physical content of the model. It is commonly accepted that $SF$ includes three terms - the three basic mechanisms of evolution: (1) the mechanism of wave-energy exchange with the wind $\mathbf{U}(t, x, y)$ ("pumping"), $In(S,U)$; (2) the mechanism of nonlinear interactions between waves ("nonlinearity"), $Nl(S)$; (3) the dissipation mechanism of wave-energy losses("dissipation"), $Dis(S,U)$.

Each of the $SF$-terms should be derived from the basic equations of the wind-wave dynamics (Komen et al, 1994; The WISE group, 2007; Polnikov, 2010a). However, even in the simplest case of ideal fluid, these equations of dynamics (e.g., the Euler equations) are not amenable to analytical solution. Nevertheless, by using a number of simplifying assumptions, methods of Fourier expansions, and perturbation theory in a small parameter, one can obtain the expected analytical forms of $SF$-terms. Though, the exact forms for $SF$-terms cannot be analytically obtained (Komen et al, 1994; The WISE group, 2007; Polnikov, 2010a). The only exception is term $Nl(S)$ (Hasselmann, 1962). Theoretical results for terms $In$ and $Dis$ (for example, the Miles' model for pumping-term (Miles, 1957), or the model for dissipation-term (Polnikov, 2013)) require a search of adequate analytical approximations and subsequent verification of them.

According to (Kitaigorodskii 1983; Zakharov and Zaslavskii, 1982; Zakharov and Badulin, 2011), the spectrum tail should be determined predominantly by the nonlinear mechanism of evolution. Herewith, in the work (Zakharov and Badulin, 2011), they even give a theoretical justification that the nonlinear mechanism should suppress the impact of other mechanisms in the equilibrium range of the wind-wave spectrum. This justification is based on an artificial sharing the exact term $Nl(S)$ into two parts: the positive part, $Nl^+$, and the negative part, $Nl^-$, (despite the strong algebra rule that similar, cubic in spectrum terms $Nl^+$ and $Nl^-$, on the contrary, are to be summed up before their comparison with the other terms). After that, Zakharov & Badulin (2011) have demonstrated that the relations $Nl^+(S) >> In(S)$ and $|Nl^-(S)| >> |Dis(S)|$ take place (when $\sigma > \sigma_p$). In their mind, that proves their general assertion.

However, our experience in wave modeling and *SF*-terms verification (Polnikov, 2005, 2010; Polnikov and Innocentini, 2008; Samiksha et al, 2015), and a number of exact calculations of *Nl(S)* (Polnikov, 1989; Polnikov and Uma, 2014) urge us to doubt that the nonlinear mechanism is the only one which is responsible for the equilibrium-tail formation in a spectrum of real wind waves. It seems that a final decision of this point can be found by the direct numerical simulations, only.

As noted above, at present there are a huge number of proven parametrizations for *In* and *Dis* (Polnikov, 2005; Komen et al, 1994; The WISE Group, 2007, among others), which allow us to solve the problem said numerically by means of varying expressions for *In* and *Dis*. Herewith, the nonlinear mechanism, *Nl(S)*, should be setup by its exact numerical representation. By this way, it is possible to clarify the role and impact of all these mechanisms on formation of the decay law for the equilibrium part of the wind-wave spectrum. If any variations of frequency dependences for *In(σ)* and *Dis(σ)* have no influence on the shape of the spectrum tail, it is obvious that the tail is fully formed by the *Nl*-mechanism. And vice versa.

The essence of this work is precisely to verify this assertion numerically.

3. **The method of research and formulation of the task**

3.1. The evolution mechanisms representations

To solve the task posed, we should specify analytical representations for the main *SF*-terms to be used. We begin with term *Nl(S)*, as the most exact and well-studied one (Polnikov, 1989, 1990, 2007).

In the frame of conservative system of Euler equations and the potential approximation, *Nl(S)* is described by the expression obtained for the first time in (Hasselmann, 1962). It has the form of a six-dimensional integral over the wave-vectors space, $\mathbf{k}(\sigma, \theta)$, which can be written as

$$Nl(S_0) \equiv \frac{\partial S(\mathbf{k}_0)}{\partial t} = 4\pi \int M^2_{0,1,2,3} S3_{0+1-2-3} \delta(\sigma_0 + \sigma_1 - \sigma_2 - \sigma_3) \delta(\mathbf{k}_0 + \mathbf{k}_1 - \mathbf{k}_2 - \mathbf{k}_3) d\mathbf{k}_1 d\mathbf{k}_2 d\mathbf{k}_3 . \quad (5)$$

The integral in (4) is called as the four-wave kinetic integral (KI). Here, $M^2_{0,1,2,3}$ are the matrix elements of the four-wave interactions; $S3_{0+1-2-3}$ is the functional of the third order in the spectrum; $\delta(\sigma)$ and $\delta(\mathbf{k})$ are the delta functions for frequency and wave vector, which means the resonant feature of the four-wave nonlinear interactions. More information about KI is not required here, as far as the integral itself is well studied. For the further, it is important only that the numerical codes for the exact calculation of KI are available to the author (Polnikov, 1989).

Regarding to choice of representation for terms *In* and *Dis*, there is a considerable arbitrariness, due to the lack of strict theories, as mentioned above. In our case, we restrict ourselves with those representation for *In* and *Dis*, which we are well known for us and have been verified in a lot of numerical simulations of real wind-wave fields (Polnikov, 2005, 2010b; Polnikov et al, 2007; Polnikov and Innocentini, 2008; Samiksha et al, 2015).

As the initial pumping function, *In(S,*$\mathbf{U}$*)*, it will be used the well-known linear in spectrum representation (Miles, 1957)

$$In = \beta(\sigma, \theta, \mathbf{U}) \sigma S(\sigma, \theta) , \quad (6a)$$

in which the growth rate, $\beta(\sigma, \theta, \mathbf{U})$, is given by the semi-empirical approach (Polnikov, 2005)

$$\beta = C_{in} \, max\left\{-b_L, \left[0.04\left(\frac{u_*\sigma}{g}\right)^2 + 0.00544\frac{u_*\sigma}{g} + 0.000055\right]cos(\theta - \theta_u) - 0.00031\right\}. \quad (6b)$$

Conversion of the wind velocity at the standard horizon, $U_{10}$, to the friction velocity, $u_*$, is carried out with the simple formula (here, accuracy of conversion is not critical)

$$u_* = U_{10}/28 . \quad (6c)$$

With no limits of generality, the value of $U_{10}$ is taken to be equal to 10 m/s, and the wind direction, $\theta_u$, is assumed to be equal to 0. With fitting constants $C_{in} = 0.5$ and $b_L = 5 \cdot 10^{-6}$ (found in Samiksha et al, 2015), the expression for *In(S,σ)*, given by forms (6a, b, c), will be considered as the "basic" one. Further variations of *In(S,σ,*U*)*, associated with a change of its frequency dependence, will differ from the basic expression by the dimensionless factor, $u_*\sigma/g$, taken to some power.

As the initial dissipation function, *Dis(S,*U*)*, it will be used the quadratic in spectrum representation proposed in (Polnikov, 2005) and theoretically justified in (Polnikov, 2012)

$$Dis(\sigma,\theta,S,\mathbf{U}) = C_{dis}c(\sigma,\theta,\sigma_p)T(\sigma,\theta,\theta_u)max[\beta_{dis},\beta(\sigma,\theta,\mathbf{U})]\frac{\sigma^6}{g^2}S^2(\sigma,\theta). \quad (7a)$$

Here, the factor $T(\sigma,\theta,\theta_u)$ is a variable element of the frequency dependence of *Dis(σ)*. In standard case, it has the form

$$T(\sigma,\theta) = 1 + T(\sigma)T(\theta) , \quad (7b)$$

where

$$T(\sigma) = 4\frac{\sigma}{\sigma_p} \quad \text{and} \quad T(\theta) = sin^2((\theta - \theta_w)/2) . \quad (7c)$$

Other details of representation (7a) are not significant here (see, for details Polnikov, 2012). With fitting constants $C_{in} = 0.55$ and $\beta_{dis} = 5 \cdot 10^{-5}$ (found in Samiksha et al, 2015), the expression for *Dis (S,σ,*U*)*, given by forms (7a, b, c), will be considered as the "basic" one. Further variations of *Dis (S,σ,*U*)*, associated with a change of its frequency dependence, will differ from the basic expression by variation of factor $T(\sigma,\theta,\theta_u)$.

3.2. Formulation of the simulation task

There are several items of the task.

1) The generalized kinetic equation of the form

$$\frac{\partial S}{\partial t} = F \equiv Nl + F_{In}(In) - F_{Di}(Dis) \quad , \tag{8}$$

taken at one space-point, is to be solved numerically (for this task, a spatial evolution is not essential). The initial form of the spectrum is given by the typical JONSWAP spectrum with an angular distribution of the form: $cos^2\theta$. As it was established long ego (Komen et al, 1994), the initial shape of the spectrum does not affect the form of solution of equation (8). This fact eliminates the need to vary the initial condition.

Solution of equation (8) is performed numerically with the algorithm proposed in (Polnikov, 1990).

2) Regarding to the *Nl*-term, it is accepted the exact calculation of KI. The *In*-term and *Dis*-term are taken in their basis forms given by formulas (6) and (7), respectively. In order to change the frequency dependences of *In* and *Dis*, the modifications said above are involved, conventionally represented in (8) by the functionals $F_{In}$ and $F_{Di}$.

3) Equation (8) is solved numerically for dimensional values of $S(\sigma,\theta)$ in the $(\sigma,\theta)$-domain defined by the boundaries [$0.6 \leq \sigma \leq 7$ rad/s; $-90^0 \leq \theta \leq 90^0$].

4) On the evolution-time scales of the order (or more) of $10^5$ of the initial peak-period, $\tau_p = 2\pi/\sigma_p$, in the range of frequencies $2\sigma_p < \sigma < 4\sigma_p$, the spectral decay-parameter *n* is determined by the least squares for the power-like spectrum shape

$$S(\sigma) = c\sigma^{-n} \quad . \tag{9}$$

5) After that, by changing functionals $F_{In}$ and $F_{Di}$, the forms of *In* and *Dis* are changed as said above, and items 3 and 4 of the task are again carried out.

6) Finally, the dependence of parameter *n* on presentations of *In* and *Dis* is determined.

7) In passing, a set of various kinds of characteristics for the spectrum evolution, $S(\sigma;t)$, is established. For example, the time-history of one-dimensional nonlinear transfer, $Nl(\sigma; t)$, and the time-history of the balance between pumping and dissipation, $B(\sigma;t)=[In(\sigma;t)-Dis(\sigma;t)]$. The latter is used for checking a presence of the inertial interval.

The main purpose of the simulations is to define the role of mechanisms *Nl*, *In* and *Dis* in formation of the spectrum-tail for wind waves and in setting the value of parameter *n*.

4. **Results and analysis**

Results of simulations are presented in Tab. 1.

Table 1.
Parameters of simulations and final values of equilibrium spectrum-tail parameter *n*

| # run | Form of $S(\sigma,\theta)_{t=0}$ | Form of *Dis* | Form of *In* | n (±5%) |
|---|---|---|---|---|
| 1 | $J(\sigma_p=2,\cos^2\theta)$ | $Dis_{base}$ | $In_{base}$ | 4.7 |
| 2 | -»- | $T(\sigma)=0.1+\dfrac{\sigma}{\sigma_p}$; $T(\theta)=1$ | $0.1*In_{base}$ | 5.1 |
| 3 | -»- | $T(\sigma)=2+8\dfrac{\sigma}{\sigma_p}$; $T(\theta)=1$ | $In_{base}$ | 5.2 |
| 4 | -»- | $T(\sigma)=2+30\dfrac{\sigma}{\sigma_p}$; $T(\theta)=1$ | $2*In_{base}$ | 5.8 |
| 5 | -»- | $Dis_{base}$ | $10\dfrac{u_*\sigma}{g}*In_{base}$ | 4.6 |
| 6 | -»- | $Dis_{base}$ | $10\left(\dfrac{u_*\sigma}{g}\right)^2*In_{base}$ | 3.8 |
| 7 | $PM(\sigma_p=2,\cos^2\theta)$ | $Dis_{base}$ | $10\left(\dfrac{u_*\sigma}{g}\right)^2*In_{base}$ | 3.8 |

Note. J means the JONSWAP shape of spectrum. PM means the Pierson-Moscovitz shape of spectrum. In the brackets, the peak frequency in r/s and angular function are shown. 5% is the mean accuracy of estimation parameter *n*.

As seen from Tab. 1, the changing form of term $Dis_{base}$ is determined only by changing functions $T(\sigma,\theta)$, $T(\sigma)$ and $T(\theta)$, defined in (7b, c); whilst term $In_{base}$ is modified by multiplication with dimensionless frequency ($u_*\sigma/g$) to the first or second power, taken with factor of 10. Meaning of these changes is a simple increase of frequency dependence for basic functions: runs 2-4 correspond to increasing frequency dependence of *Dis(σ)*, whilst runs to increasing frequency dependence of *In(σ)*.

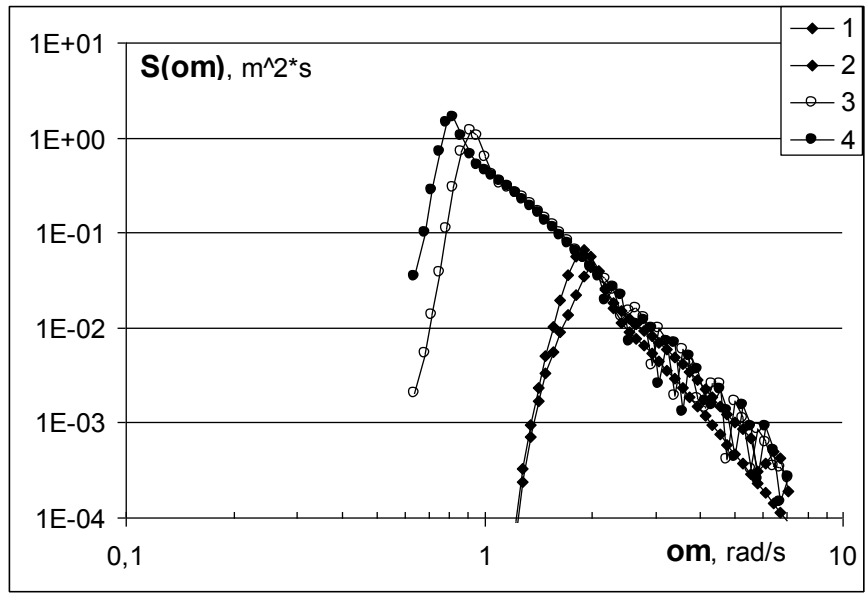

Fig. 1a. Evolution of one-dimensional spectrum *S(σ)* for run 1. Line 1 corresponds to time $t=0$, line 2 to $t = 1.2 \cdot 10^3$s, line 3 to $t = 8.8 \cdot 10^4$s, line 4 to $t = 1.8 \cdot 10^5$s.

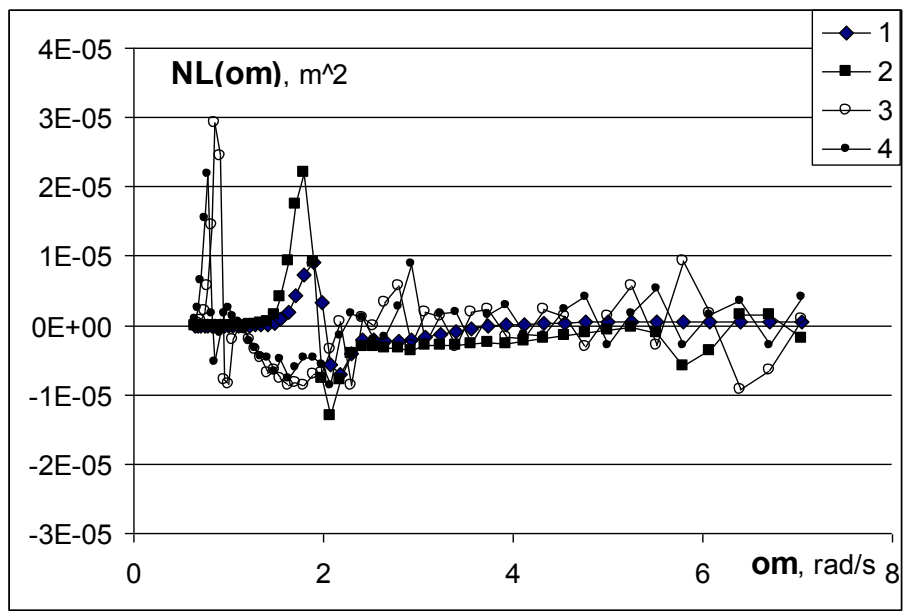

Fig. 1b. Evolution of one-dimensional nonlinear term *Nl(σ)* for run 1. For legend see Fig.1a.

It is seen that the increase of frequency dependence *Dis(σ)* leads to growing the power of decay for equilibrium range of the wind-wave spectrum. A similar strengthening of frequency dependence *In(σ)* results in a going down the power of decay of the spectrum.

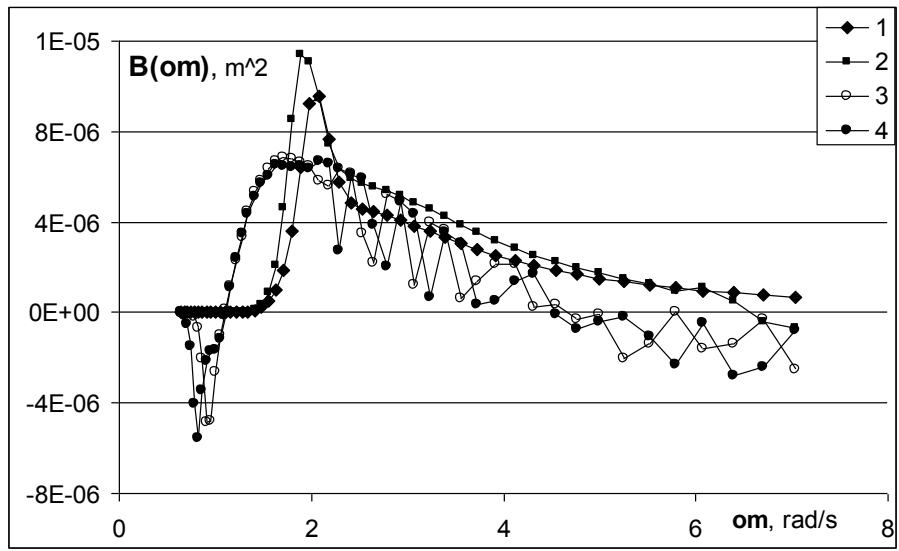

Fig. 1c. Evolution of one-dimensional balance $B(\sigma)$ for run 1. For legend see Fig.1a.

In Figs 1, 2, 3, some results for $S(\sigma;t)$, $Nl(\sigma;t)$, and $B(\sigma;t)$ are given in the graphical form. As was expected, it is perfectly seen from Figs 1a, 2a, 3a a setting the equilibrium shapes of the spectrum in the drop-down tail, having insignificant oscillations of intensity. The spectrum fluctuations at higher frequencies, as known (Polnikov, 1990), are due to numerical errors in calculations of $Nl(\sigma;t)$ (Figs 1b, 3b.). It is provided by setting fixed values of time-steps $\Delta t$ during numerical solution of (8) for all spectral components, simultaneously. Usually such fluctuations are suppressed numerically by their smoothing. However, in view of satisfactory shape of the simulation output, in this study the issue of smoothing solutions was not elaborated upon in details.

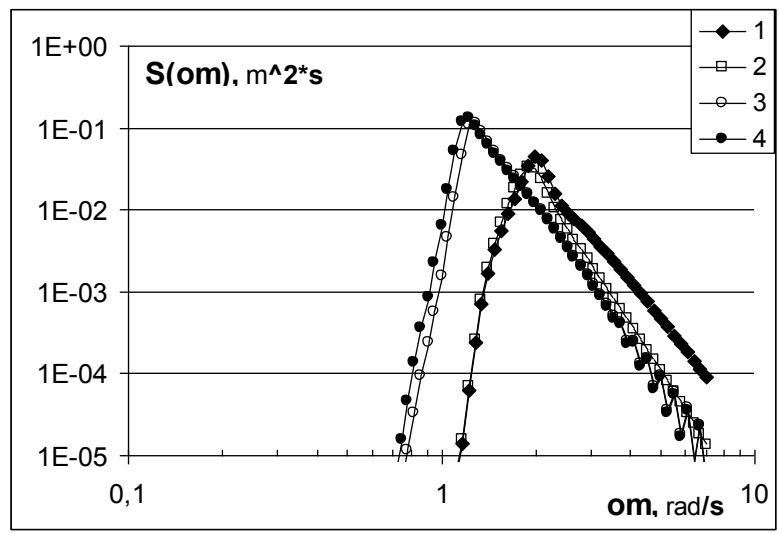

Fig. 2a. . Evolution of one-dimensional spectrum $S(\sigma)$ for run 4. Line 1 corresponds to time $t=0$, line 2 to $t=8.2\cdot 10^2$s, line 3 to $t=1.0\cdot 10^5$s, line 4 to $t=1.4\cdot 10^5$s.

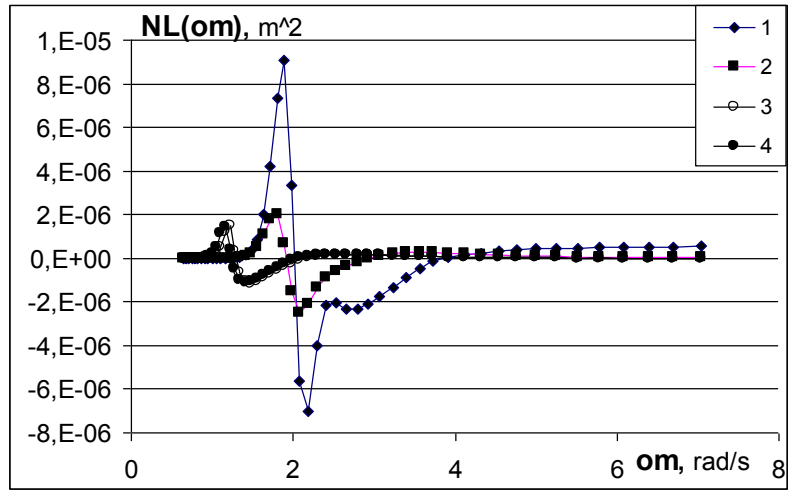

Fig. 2b. Evolution of one-dimensional nonlinear term $Nl(\sigma)$ for run 4. For legend see Fig.2a.

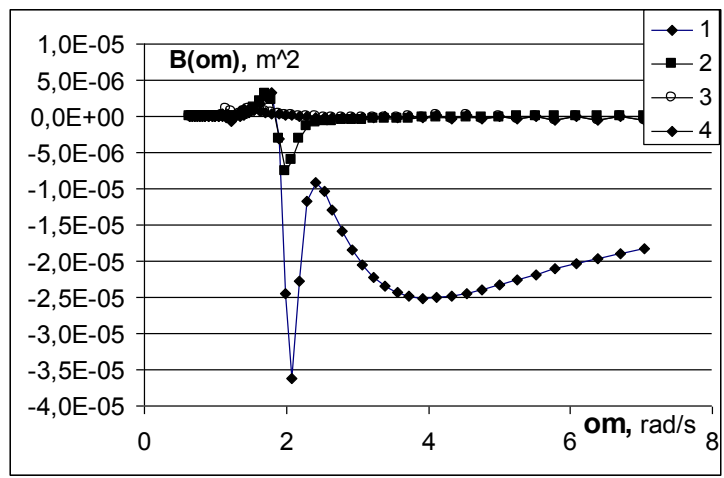

Fig 2c. Evolution of one-dimensional balance $B(\sigma)$ for run 4. For legend see Fig.2a.

In view of the task posed, the analysis of time-history for nonlinear mechanism $Nl(\sigma; t)$ has no physical interest. From Figs. 1b, 2b, and 3b, it is seen the regular moving of the remarkable nonlinear transfer (linked to the current spectral peak) to the lower frequencies, accompanied with a decreasing of the nonlinear transfer intensities. In the equilibrium range of spectrum, the time-averaged value of $Nl(\sigma; t)$ is very small.

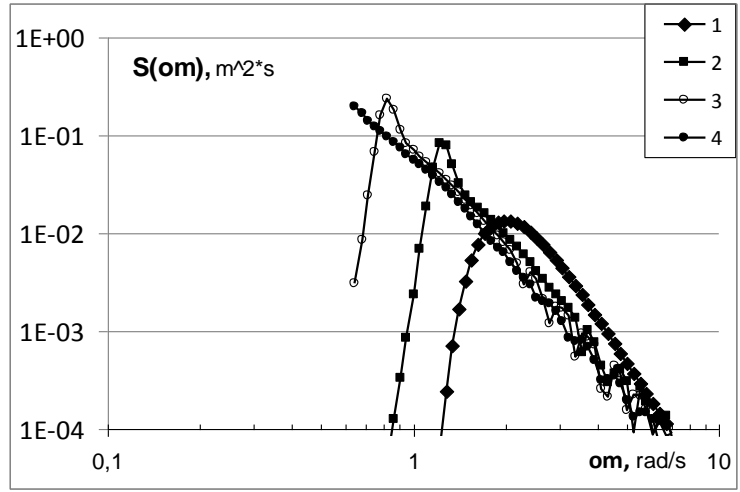

Fig. 3a. Evolution of one-dimensional spectrum $S(\sigma)$ for run 7. Line 1 corresponds to time $t=0$, line 2 to $t = 4.3 \cdot 10^3$s, line 3 to $t =9.0 \cdot 10^5$s, line 4 to $t =2.0 \cdot 10^6$s.

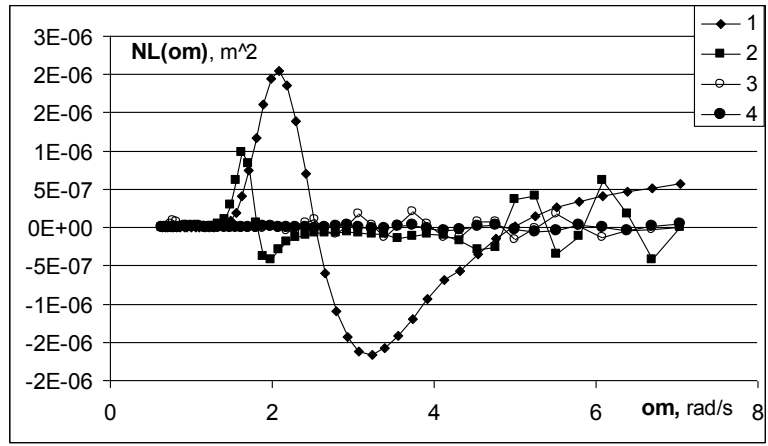

Fig. 3b. Evolution of one-dimensional nonlinear term $Nl(\sigma)$ for run 7. For legend see Fig.3a.

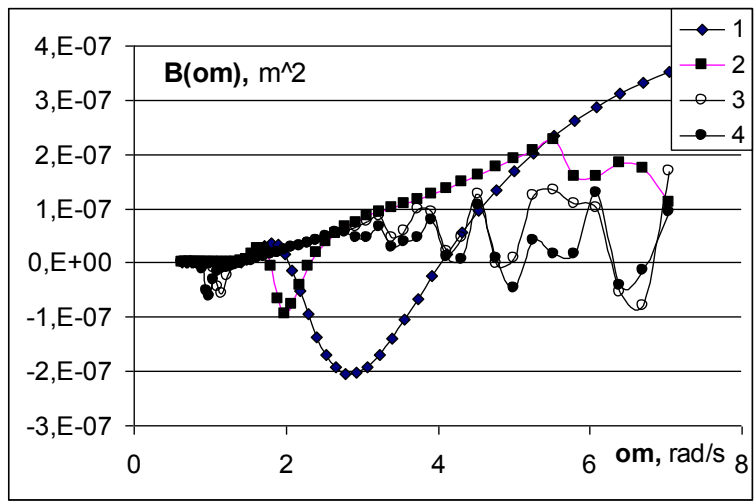

Fig. 3c. Evolution of one-dimensional balance $B(\sigma)$ for run 7. For legend see Fig.3a.

Analysis of the time-history for balance $B(\sigma;t)$ is much more interesting. First of all, it is seen from Fig.1c that, in the case of real wind-wave conditions (basic *In* and *Dis*), there is no any inertial frequency interval. From the very beginning (curves 1 and 2) corresponding to evolution time $t \leq 100\ \tau_p$, real balance is positive everywhere in the considered frequency band. Thus, there is no condition for the Kolmogorov's spectra formation.

Second, as shown in Figs.1c, 2c, 3c, during evolution for more than several hundreds of periods $\tau_p$, at high frequencies ($\sigma > 2\sigma_p$) balance $B(\sigma;t)$ becomes to be oscillating around a zero value. It means that the time-averaged value of balance is close to zero, i.e.

$$< (In(\sigma;t) - Dis(\sigma;t)) > \approx 0 \qquad (\sigma > 2\sigma_p). \tag{10}$$

It is this balance of pumping and dissipation, taking place even on the background of fluctuations of $Nl(\sigma; t)$, gives a hint for explanation of the results.

Indeed. It is naturally to assume that the equilibrium range of the spectrum $S(\sigma)$ is formed due to the balance of pumping and dissipation, which is being realized for a certain, equilibrium form of spectrum $S_{eq}(\sigma)$, i.e. when the spectrum $S(\sigma)$ is equal to $S_{eq}(\sigma)$. Then, writing an analytical presentations of terms *In* and *Dis* in the forms

$$In = \alpha\ (u^*,\sigma)S(\sigma), \qquad \text{and} \quad Dis = \gamma\ (u^*,\sigma)\ S^2(\sigma), \tag{11}$$

from condition of balance (10), taking place at for the spectrum $S(\sigma) = S_{eq}(\sigma)$, we immediately obtain for the tail range of frequencies ($\sigma > 2\sigma_p$) an expected shape of the equilibrium spectrum of the form

$$S_{eq}(u_*,\sigma) \approx \alpha(u_*,\sigma)/\gamma\ (u_*,\sigma). \tag{12}$$

The results presented in Tab. 1 confirm clearly the said. Moreover, according to ratio (12), the equilibrium range of wind-wave spectrum should not obligatory be exactly of power-like form (9).

Hence, by a reasonable varying the forms of $In(u_*,\sigma)$ and $Dis(u_*,\sigma)$, in the numerical solution of equation (8) one can obtain any, preassigned form of $S_{eq}(u_*,\sigma)$. Therefore, the above results directly confirm the said interpretation of the tail formation for the wind-wave spectrum. That finally closes the task.

## 5. Conclusions and their discussion

The first and most important conclusion to be drawn from the results presented above is the following. Only the balance between the input mechanism, $In(\sigma)$, and the dissipation mechanism, $Dis(\sigma)$, provides the equilibrium range for the wind-wave spectrum. Herewith, we state that this balance is the most feasible if the dissipation term is of the second (or higher) power in spectrum (as it was justified analytically in Polnikov, 2012). Otherwise, there is no reason for the equilibrium spectrum formation in real wind waves, as far as the total nonlinear term, $Nl(\sigma)$ is fairly small at high frequencies with respect to balance $B(\sigma) = In(\sigma) - Dis(\sigma)$.

Second. In the case of basic forms of $In(\sigma)$ and $Dis(\sigma)$, usually used for simulation of real wind waves (Polnikov, 2005, 2010; Samiksha et al, 2015), there is no visible inertial interval in the frequency band. Thus, here is no reason for the Kolmogorov-type formation of the equilibrium spectrum tail.

Third. It is clear that quite different conditions of wave evolution could be realized in the nature. In this case, the actual frequency dependences of pumping and dissipation functions could also vary, differing from the usually used, standard (or basic) ones. Therefore, according to (11, 12), in the nature one may observe a wide range of values for the falling-law parameter, $n$, corresponding to the equilibrium range of the wind-wave spectrum. This gives a clear interpretation of the results obtained in (Rodrigues & Soares, 1999).

In addition, formulas (11, 12) allow us to give an answer to the question: why the spectra of form (2) are often observed in water reservoirs of small sizes (laboratory tanks, lakes, etc.)? Apparently, this is due to a small temporal evolution of these spectra, when dissipative processes have some specific dynamics, provided by, for example, specificity of wave breaking and intercity of the upper layer mixing. According to the theory (Polnikov, 2012), this can result in appearing the dissipative term in a form that differs from one for large water areas. For example, the specificity of breaking in water reservoirs of small sizes can lead to a weakening the frequency dependence *Dis(σ)*, and result in a weaker law of spectra decay, observed in laboratory tanks (Toba, 1972) or in lakes (Donelan et al, 1985).

It is also possible that at a short evolution-time, the decay law "-4" is formed due to the pumping, having the "white noise" spectrum, as described in (Polnikov & Uma, 2014).

Thus, in any case, it is the balance of pumping and dissipation is responsible for the equilibrium range formation of the wind-wave spectrum. The role of the nonlinear mechanism is limited to the wave enlargement, not having a noticeable effect on the spectrum-tail shape. Under idealized conditions, only, when a fairly wide inertial frequency interval is artificially implemented, the flux spectra of Kolmogorov-type (2, 3) could be realized, as was shown numerically in (Polnikov, 1994, 2001).

**Acknowledgments**

The author thanks for the remarks of participants of the nonlinear session at the Shirshov Institute of Oceanology of RAS (December, 2016), where this work was presented. The work was supported by a grant of the Russian Science Foundation, #14-27-00134.